\documentclass[10pt, conference, letterpaper]{IEEEtran}
\usepackage{amsmath}
\usepackage{amsthm}
\usepackage{amssymb}
\usepackage{ifthen}
\usepackage{graphicx}
\usepackage{subfigure}
\usepackage{color}
\usepackage{balance}
\usepackage{cite}
\usepackage{url}
\begin{document}
\title{A Min-plus Model of Age-of-Information \\ with Worst-case and Statistical Bounds}
\author{\IEEEauthorblockN{Mahsa Noroozi, Markus Fidler} 
\IEEEauthorblockA{Institute of Communications Technology \\ Leibniz Universit\"{a}t Hannover}
}
\maketitle
\thispagestyle{plain} 
\pagestyle{plain}
\begin{abstract}
We consider networked sources that generate update messages with a defined rate and we investigate the age of that information at the receiver. Typical applications are in cyber-physical systems that depend on timely sensor updates. We phrase the age of information in the min-plus algebra of the network calculus. This facilitates a variety of models including wireless channels and schedulers with random cross-traffic, as well as sources with periodic and random updates, respectively. We show how the age of information depends on the network service where, e.g., outages of a wireless channel cause delays. Further, our analytical expressions show two regimes depending on the update rate, where the age of information is either dominated by congestive delays or by idle waiting. We find that the optimal update rate strikes a balance between these two effects.
\end{abstract}
%
%
\section{Introduction}
\label{sec:introduction}
Networked cyber-physical systems rely on timely status information provided by all kinds of remote sensors. At a data sink, the freshness of a status information depends on network delays but also on the update rate of the sensor. Age of information quantifies this freshness, measuring the time that has elapsed between the generation of a sensor reading and its use.

\begin{figure}
\centering
\includegraphics[width=\linewidth]{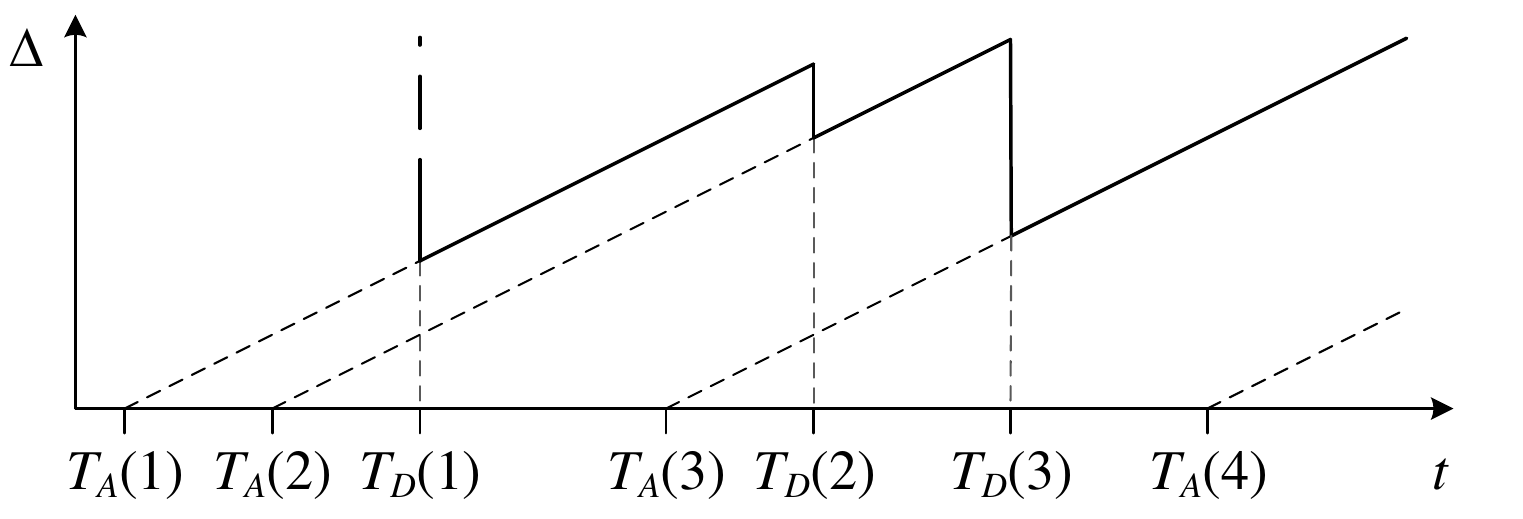}
\caption{Progression of age of information $\Delta(t)$ over time $t$. $T_A(n)$ and $T_D(n)$ are the arrival and departure time stamps of status update $n$.}
\label{fig:aoisimple}
\end{figure}

A common illustration~\cite{kaul:ageofinformationvehicular} of the progression of the age of information $\Delta(t)$ over time $t \ge 0$ is shown in Fig.~\ref{fig:aoisimple}, where $T_A(n)$ denotes the arrival time of status update $n \ge 1$ from the sensor to the network and $T_D(n)$ its departure time from the network to the sink. The slope of the upward segments is one. For an example, select a time $t^* \ge T_D(1)$, determine the latest status update at the sink $n^* = \max \{n \ge 1: T_D(n) \le t^*\}$ that was generated at time $T_A(n^*)$ to find the age of information $\Delta(t^*) = t^* - T_A(n^*)$.

The notion of age of information has been introduced in vehicular networks~\cite{kaul:ageofinformationvehicular} where it has been referred to also as status age~\cite{kaul:ageofinformationqueue}, information freshness~\cite{zinchenko:informationfreshness}, or message lifetime~\cite{tchouankem:messagelifetime}. It emerged as a very active area of research, being of general importance for a variety of applications in the areas of cyber-physical systems and the Internet of Things. Particular challenges arise in networked feedback control systems~\cite{champati:ageofinformationfeedbackcontrol,champati:ageofinformationmaxplus}. Further applications extend to cache updating and microblogging~\cite{altmann:ageofinformationmicroblogging}. For a recent, comprehensive survey see~\cite{yates:ageofinformationsurvey}.

A focus of age of information research are wireless channels, such as multiple access channels~\cite{kaul:ageofinformationvehicular,pappas:ageofinformationnetworkcalculus}, memoryless on-off channels~\cite{champati:ageofinformationmaxplus}, fading channels~\cite{pappas:ageofinformationnetworkcalculus}, and wireless networks with interference constraints~\cite{modiano:informationfreshness}. With the general aim of minimizing the age of information, a frequent subject of investigation is the optimal update rate~\cite{kaul:ageofinformationqueue,modiano:ageofinformationqueueing,champati:ageofinformationmaxplus,champati:ageofinformationdgqueue}. Typical update processes are periodic~\cite{kaul:ageofinformationvehicular, champati:ageofinformationmaxplus, champati:ageofinformationdgqueue} or random~\cite{kam:ageofinformation, pappas:ageofinformationnetworkcalculus, modiano:ageofinformationqueueing, modiano:informationfreshness, champati:ageofinformationfeedbackcontrol, kaul:ageofinformationqueue}. Predominantly, these works use analytical models, such as queueing theory~\cite{kaul:ageofinformationqueue, modiano:ageofinformationqueueing, champati:ageofinformationgigiqueue}, to derive time averages of mean and peak age of information.

Most closely related to this work are two papers that apply techniques common to the network calculus:~\cite{pappas:ageofinformationnetworkcalculus} derives statistical delay bounds using a $(\min, \times)$-algebra with Mellin transform and moment bound; and~\cite{champati:ageofinformationmaxplus} employs a max-plus algebra to derive statistical age of information bounds by Chernoff's bound. The max-plus formulation works with time stamps of packet arrivals and departures. It can express delay most easily and extends naturally to age of information.

Differently from~\cite{champati:ageofinformationmaxplus}, we choose a min-plus algebra in this work. The min-plus approach uses cumulative arrivals and departures, i.e., bits as functions of time, that are pseudo-inverse functions of the max-plus representation~\cite{liebeherr:duality}. Unlike max-plus, min-plus can conveniently model multiplexing of traffic flows and time-varying services that arise, e.g., due to wireless communications or scheduling algorithms. We will take advantage of this and derive statistical age of information bounds for these systems. Our results show how service outages, congestive delays, and idle waiting affect the age of information. They enable finding the optimal update rate.

The remainder of this paper is structured as follows. We will derive our min-plus model of the age of information in Sec.~\ref{sec:minplusmodel}, where we also show worst-case bounds for periodic updates. In Sec.~\ref{sec:randomservice} we derive a solution for random service and show statistical age of information bounds for a Markov channel. Random arrivals are considered in Sec.~\ref{sec:randomarrivals} and scheduling of multiple sources with different priorities in Sec.~\ref{sec:scheduling}. We present brief conclusions in Sec.~\ref{sec:conclusions}.

%
%
\section{Min-Plus AoI Model}
\label{sec:minplusmodel}
We consider a queueing system, such as a buffered link, a scheduler, or a network thereof. We denote arrivals $A(t)$ the cumulative amount of data arriving at the system in $[0,t)$. By definition, the function $A(t)$ is non-negative, non-decreasing and passes through the origin. For convenience we extend the definition $A(t) = 0$ for $t < 0$. Hence, $A(t) \in \mathcal{F}_0$ where \mbox{$\mathcal{F}_0 = \{f(t): f(t) \ge f(\tau) \ge 0 \,\, \forall t \ge \tau \ge 0, f(t) = 0 \,\, \forall t \le 0 \}$}. We use shorthand notation $A(\tau,t) = A(t) - A(\tau)$ for $t \ge \tau$. We will use a continuous time model $t \in \mathbb{R}$ where by convention $A(t)$ is left-continuous. Similarly, the cumulative departures of a system are denoted $D(t) \in \mathcal{F}_0$, where in addition $D(t) \le A(t)$ for all $t$ for causality.

We use the concept of dynamic server~\cite{chang:dynamicserviceguarantees} to model systems. \mbox{A system has service process $S(\tau,t)$ for $t \ge \tau \ge 0$ if}
\begin{equation}
D(t) \ge \inf_{\tau \in [0,t]} \{ A(\tau) + S(\tau,t) \} =: A \otimes S(t),
\label{eq:serviceprocess}
\end{equation}
for all $t \ge 0$, where $S(\tau,t)$ is non-negative, non-decreasing with $t$, and non-increasing with $\tau$. For a basic example, a buffered, lossless, and work-conserving link with constant service rate $c > 0$ has service process $S(\tau,t) = c (t-\tau)$. Another example is a link with a time-varying service rate~\cite{chang:performanceguarantees}. We will use stochastic service processes $S(\tau,t)$ in Sec.~\ref{sec:randomservice}.

In a continuous time model, it is convenient to assume that data behaves like fluid, i.e., a system may serve any amount of data regardless of packet or message boundaries. The effects that are due to packet boundaries are modeled by a packetizer that converts fluid input $x \in \mathbb{R}_+$ to packetized output $P_L(x)$~\cite{leboudec:networkcalculus, chang:performanceguarantees}. Given packets of length $l(n) > 0$ with packet index $n \in \mathbb{N}$, we denote the cumulative packet length $L(n) = \sum_{\nu=1}^n l(n)$ and $L(0) = 0$. \mbox{The output of the packetizer is}
\begin{equation}
P_L(x) = \max_{n \in \mathbb{N}_0} \bigl\{ L(n) 1_{\{L(n) \le x\}} \bigr\} ,
\label{eq:packetizer}
\end{equation}
where the indicator function $1_{\{.\}}$ is one if the argument is true and zero otherwise. It follows that \mbox{$x \ge P_L(x) \ge x -l_{\max}$,} where $l_{\max} = \max_{n \in \mathbb{N}} \{l(n)\}$ is the maximal packet length. A function $A(t)$ is packetized if $A(t)=P_L(A(t))$.

Fig.~\ref{fig:cumulativefunctions} shows an example of a packetized arrival function $A(t)$ (left-continuous, marked by empty and full circles) and departure function $D(t)$, fluid (dashed line) and packetized (solid line), respectively.

Now, consider the series of a fluid system and a packetizer. The system has service process $S(\tau,t)$, packetized arrivals $A(t) = P_L(A(t))$, and fluid departures $D(t) \ge A \otimes S(t)$. For the packetized departures $P_L(D(t))$ it is known that~\cite{leboudec:networkcalculus, chang:performanceguarantees}
\begin{align}
P_L(D(t)) & \ge \inf_{\tau \in [0,t]} \{ P_L (A(\tau) + S(\tau,t)) \} \nonumber \\
& \ge \inf_{\tau \in [0,t]} \{ A(\tau) + [S(\tau,t)-l_{\max}]_+ \},
\label{eq:servicecurvepacketizer}
\end{align}
where $[x]_+ = \max \{0,x\}$. The first line follows since $P_L(x)$ is non-decreasing and right-continuous. In the second line, $P_L(x) \ge x - l_{\max}$ and $P_L (A(\tau)+S(\tau,t)) \ge P_L(A(\tau)) = A(\tau)$ since $S(\tau,t)$ non-negative and $A(t)$ packetized are used. As a result, the effects that are due to packetization are integrated into the service process, i.e., for $t \ge  \tau \ge 0$ the combination of the fluid system and the packetizer offers service process
\begin{equation}
S_{P_L}(\tau,t) = [S(\tau,t) - l_{\max}]_+ .
\label{eq:servicepacketizer}
\end{equation}
%
%
\subsection{Definition of Age of Information}
\begin{figure}
\centering
\includegraphics[width=\linewidth]{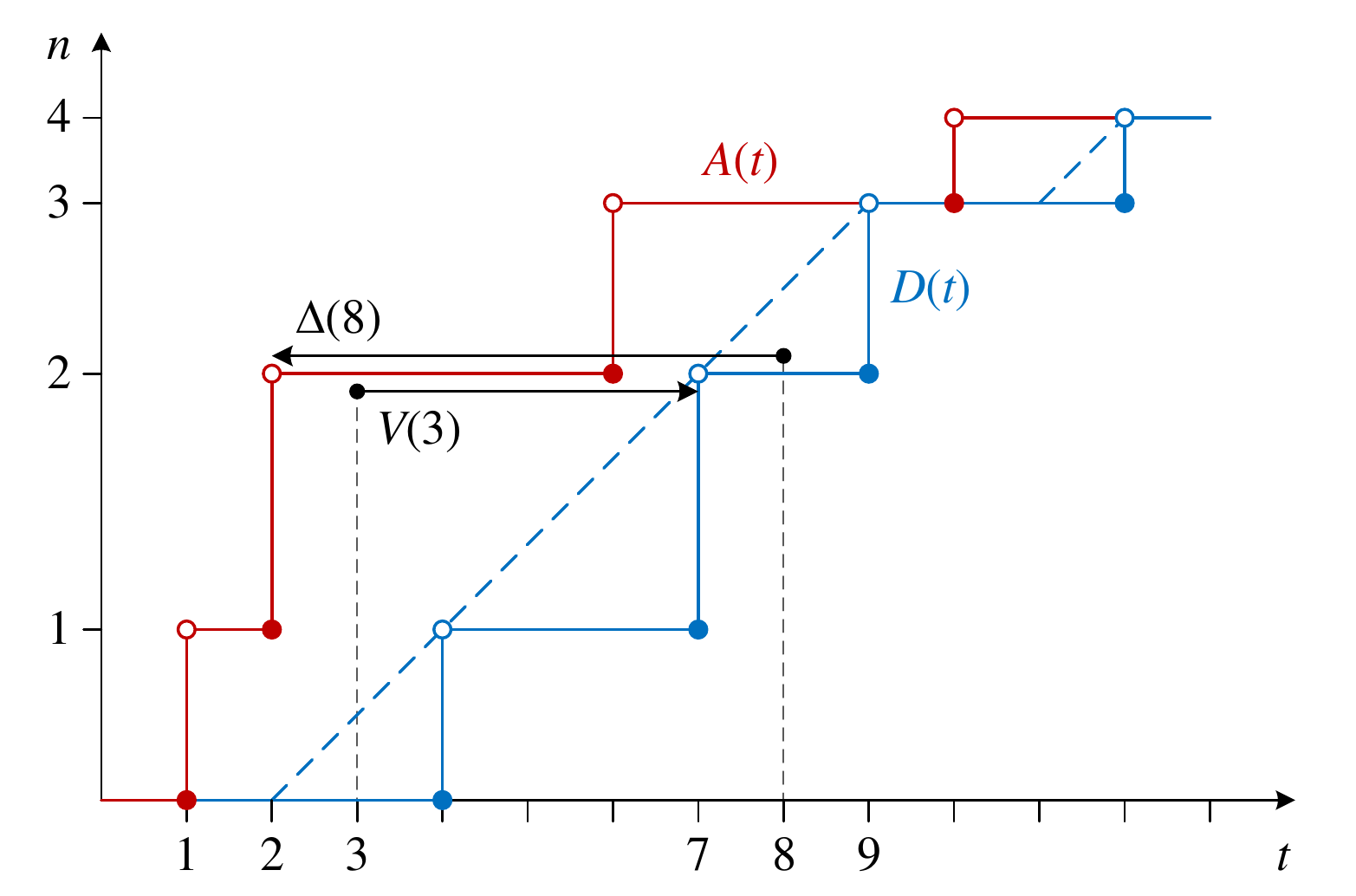}
\caption{Cumulative arrivals $A(t)$ (packetized, left-continuous) and departures $D(t)$ (dashed line fluid model, solid line packetized model) of a system, including examples of age of information $\Delta(t)$ and virtual delay $V(t)$.}
\label{fig:cumulativefunctions}
\end{figure}

Equipped with basics of the network calculus, we now derive the age of information $\Delta(t)$ at time $t \ge 0$. For a system with first-come first-served (fcfs) policy, we define
\begin{equation}
\Delta(t) = \sup \{ \delta \in [0,t] : D(t) - A(t-\delta) \le 0 \} ,
\label{eq:aoi_def}
\end{equation}
i.e., the last bit that departed before or at $t$ arrived no earlier than $t-\Delta(t)$. Hence, $\Delta(t)$ is the age of that bit in the system at time $t$. For an example, Fig.~\ref{fig:cumulativefunctions} shows $\Delta(t)=6$ for $t=8$.

Most closely related works~\cite{champati:ageofinformationgigiqueue,champati:ageofinformationmaxplus} use a max-plus model where $T_A(n)$ and $T_D(n)$ denote the arrival and departure time stamps of packet $n \in \mathbb{N}$. There is a duality of max-plus and min-plus models, see~\cite{liebeherr:duality}. The max-plus definition of age of information is $\Delta(t) = t - \max_{n \ge 1} \{T_A(n) : T_D(n) \le t\}$~\cite{champati:ageofinformationgigiqueue,champati:ageofinformationmaxplus}. The equivalence with~\eqref{eq:aoi_def} is seen in Fig.~\ref{fig:cumulativefunctions}, exemplified for $t=8$ and $n=2$.

Next, we derive the age of information for a system with service process $S(\tau,t)$. By insertion of~\eqref{eq:serviceprocess} into~\eqref{eq:aoi_def} it follows for $t \ge 0$ that
\begin{equation*}
\Delta(t) \le \sup \Bigl\{ \delta \in [0,t] : \inf_{\tau \in [0,t]} \{ S(\tau,t) + A(\tau) - A(t-\delta) \}  \le 0 \Bigr\} ,
\end{equation*}
which is equivalent to
\begin{multline}
\Delta(t) \le \sup \biggl\{ \delta \in [0,t] :  \inf \Bigl \{ \inf_{\tau \in [0,t-\delta]} \{ S(\tau,t) - A(\tau,t-\delta) \} , \\
\inf_{\tau \in (t-\delta,t]} \{ S(\tau,t) + A(t-\delta,\tau) \} \Bigr \} \le 0 \biggr\} .
\label{eq:aoi}
\end{multline}
Here, we have to pay attention to the second line of~\eqref{eq:aoi}, where $\inf_{\tau \in (t-\delta,t]} \{ S(\tau,t) + A(t-\delta,\tau) \} \ge 0$ trivially, but may nevertheless attain the outer infimum if the functions have plateaus, i.e., if they are not strictly increasing.
%
%
\subsection{Relation with Delay}
Given that age of information is a delay-based metric, we clarify its relation with the common notion of delay. In network calculus, the virtual delay for a fcfs system at time $t \ge 0$ is defined as~\cite[Def. 1.1.1]{leboudec:networkcalculus}
\begin{equation}
V(t) = \inf \{\upsilon \ge 0 : D(t+\upsilon) - A(t) \ge 0 \},
\label{eq:vd_def}
\end{equation}
i.e., the last bit that arrived before or at $t$ departed no later than $t+V(t)$. The definition of delay is virtual in the sense that it is not conditioned on an actual arrival at $t$. Fig.~\ref{fig:cumulativefunctions} shows $V(t)$ for $t=3$ as an example.

By insertion of~\eqref{eq:serviceprocess} into~\eqref{eq:vd_def} it follows for $t \ge 0$ that
\begin{equation}
V(t) \le \inf \Bigl\{ \upsilon \ge 0 : \inf_{\tau \in [0,t]} \{ S(\tau,t+\upsilon) - A(\tau,t) \} \ge 0 \Bigr\} .
\label{eq:vd}
\end{equation}
Note that by insertion of~\eqref{eq:serviceprocess} for $D(t+\upsilon)$, \eqref{eq:vd} actually considers $\tau \in [0,t+\upsilon]$. Since $\inf_{\tau \in [t,t+\upsilon]} \{ S(\tau,t+\upsilon) + A(\tau) - A(t) \} \ge 0$ generally, we can omit $\tau \in (t,t+\upsilon]$. This is different from the case of age of information~\eqref{eq:aoi}.

To make a connection of age of information with virtual delay, we fix $t \ge 0$ and define the function
\begin{equation*}
F(\delta) = D(t) - A(t-\delta) ,
\end{equation*}
that is the inner part of the age of information expression~\eqref{eq:aoi_def}. By variable substitution $t' = t-\delta$ we immediately obtain the inner part $F(\delta) = D(t'+\delta) - A(t')$ of the virtual delay~\eqref{eq:vd_def}. We can write the age of information~\eqref{eq:aoi_def} as $F^{\uparrow}(0)$, where
\begin{equation*}
F^{\uparrow}(\varphi) = \sup \{\delta : F(\delta) \le \varphi \} = \inf \{\delta : F(\delta) > \varphi \} ,
\end{equation*}
is known as the upper pseudo-inverse of $F(\delta)$. The second statement holds if $F(\delta)$ is non-decreasing, see~\cite[Sec. 10.1]{liebeherr:duality} for details on pseudo-inverse functions and their use in the network calculus. We note that $F(\delta)$ is non-decreasing by definition since $A(t)$ is non-decreasing. The virtual delay~\eqref{eq:vd_def}, on the other hand, equals the lower pseudo-inverse $F^{\downarrow}(0)$ defined as
\begin{equation*}
F^{\downarrow}(\varphi) = \inf \{\delta : F(\delta) \ge \varphi \} .
\end{equation*}
It holds that $F^{\uparrow} \ge F^{\downarrow}$ and if $F$ is continuous and strictly increasing then $F^{\uparrow} = F^{\downarrow} = F^{-1}$~\cite{liebeherr:duality}. It follows that the age of information at time $t$ equals the virtual delay at time $t'$ if $F$ is continuous and strictly increasing, whereas in case of non-decreasing, discontinuous packet arrival functions the age of information is generally not smaller than the virtual delay.
%
%
\subsection{Worst-Case Analysis}
\label{sec:worstcase}
The network calculus uses deterministic envelope functions for worst-case analysis. An upper envelope $\overline{\mathcal{E}}(t) \in \mathcal{F}_0$ and a lower envelope $\underline{\mathcal{E}}(t) \in \mathcal{F}_0$ of the arrivals satisfy for all $t \ge \tau \ge 0$ that
\begin{equation}
\overline{\mathcal{E}}(t-\tau) \ge A(\tau,t) \ge \underline{\mathcal{E}}(t-\tau).
\label{eq:envelope}
\end{equation}
Also, a deterministic lower envelope $\mathcal{S}(t) \in \mathcal{F}_0$  of the service satisfies for all $t \ge \tau \ge 0$ that
\begin{equation}
S(\tau,t) \ge \mathcal{S}(t-\tau).
\label{eq:servicecurve}
\end{equation}
By insertion of~\eqref{eq:servicecurve} into~\eqref{eq:serviceprocess} it follows that $D(t) \ge  A \otimes \mathcal{S}(t)$ $= \inf_{\tau \in [0,t]} \{ A(\tau) + \mathcal{S}(t-\tau) \}$ where $\mathcal{S}(t)$ is known as deterministic lower service curve~\cite{leboudec:networkcalculus,chang:performanceguarantees}. By insertion of~\eqref{eq:envelope} and~\eqref{eq:servicecurve} into~\eqref{eq:aoi}, a variable substitution, and letting $t \rightarrow \infty$, we find the worst-case age of information bound
\begin{multline}
\Delta_{\max} \le \sup \biggl\{ \delta \ge 0 :  \inf \Bigl \{ \inf_{\tau \ge \delta} \{ \mathcal{S}(\tau) - \overline{\mathcal{E}}(\tau-\delta) \} , \\
\inf_{\tau \in [0,\delta)} \{ \mathcal{S}(\tau) + \underline{\mathcal{E}}(\delta-\tau) \} \Bigr \} \le 0 \biggr\} .
\label{eq:aoi_det}
\end{multline}
The derivation of the age of information bound is similar to that of the worst-case virtual delay bound~\cite{leboudec:networkcalculus,chang:performanceguarantees} that follows by substitution of $\overline{\mathcal{E}}(t)$ defined in~\eqref{eq:envelope} and $\mathcal{S}(t)$ in~\eqref{eq:servicecurve} into~\eqref{eq:vd}. A major difference is that the age of information bound depends on the lower envelope function of the arrivals whereas the virtual delay does not.

Another difference between the worst-case age of information and the worst-case virtual delay arises with respect to packetization. Generally,~\eqref{eq:servicepacketizer} includes the packetizer into the service process. It is, however, known that the packetizer does not affect the worst-case virtual delay bound that is obtained for the fluid system. A proof can be found in~\cite{leboudec:networkcalculus} while Fig.~\ref{fig:cumulativefunctions} can provide some intuition: the maximal virtual delay is attained for packet $n=2$ and is determined as the maximal horizontal deviation of arrivals and departures, that is $V_{\max} = 5$, both in case of the fluid departure function as well as in case of the packetized departures. A similar result can, however, not be obtained for the age of information, and Fig.~\ref{fig:cumulativefunctions} provides a counter-example: Clearly, the maximal age of information is $\Delta_{\max} = 7$ at $t=9$ since packet $n=2$ that arrives at $t=2$ is eventually replaced by packet $n=3$ that departs at $t=9$, whereas the maximal age of information obtained for the fluid departure function is $\Delta_{\max} = 5$ obtained for $t=7$.
%
%
\subsection{Periodic Updates}
\label{sec:periodicupdates}
We consider a source that periodically sends update messages resulting in packets of length $l > 0$. Denote the width of the update interval $w > 0$, hence
$A(t) = l \, \lceil t/w \rceil$, and $A(t)$ has the upper and lower envelope functions
\begin{equation*}
l \, \lceil t/w \rceil  \ge A(\tau,\tau+t) \ge  l \, \lfloor t/w \rfloor ,
\end{equation*}
for all $\tau,t \ge 0$. For a first example, we use a fluid service curve $\mathcal{S}(t) = c t$ that models a minimal capacity guarantee $c$. Including the packetizer into the service curve model~\eqref{eq:servicepacketizer}, we obtain with $l_{\max} = l$ that $\mathcal{S}_{P_L}(t) = [c t - l]_+ = c [t-l/c]_+$. Here, packetization is expressed as a latency $t_0 = l/c$, so that the resulting service curve is of the latency-rate type $\mathcal{S}_{P_L}(t) = c [t-t_0]_+$. By insertion into~\eqref{eq:aoi_det} we have
\begin{multline*}
\Delta_{\max} \le \sup \biggl\{ \delta \ge 0 :
\inf \Bigl \{  \inf_{\tau \ge \delta} \{ c [\tau-t_0]_+ - l \, \lceil (\tau-\delta)/w \rceil \} , \\
\inf_{\tau \in [0, \delta)} \{ c [\tau-t_0]_+ + l \, \lfloor (\delta-\tau)/w \rfloor \} \Bigr \} \le 0 \biggr\} .
\end{multline*}
With the stability condition $c \ge l/w$ we find that the first infimum is smaller or equal zero only if $\delta < l/c + t_0$, consider $\tau = l/c + t_0$, and larger than zero otherwise. The second infimum is zero only if $\delta < w + t_0$, consider $\tau = t_0$, so that
\begin{equation*}
\Delta_{\max} \le \max \biggl\{\frac{l}{c} , w \biggr\} + t_0 ,
\end{equation*}
and since $w \ge l/c$ from the stability condition, we have
\begin{align}
\Delta_{\max} \le & w + t_0  \label{eq:lraoi} \\
= & w + \frac{l}{c} . \nonumber
\end{align}

We can easily see how the bound is attained if the system transmits messages with rate $c$, i.e., message $n+1$ is generated $w$ units of time after message $n$ and requires $l/c$ units of time for transmission until it can replace message $n$ at the receiver. Clearly, for a system with deterministic service curve, the maximal age of information can be reduced by decreasing the width of the update interval up to $w = l/c$, i.e., full utilization, so that $\Delta_{\max} \le 2 l/c$. In the next sections, we will use stochastic arrival and service models that will make finding the optimal update rate more tricky.
%
%
\subsection{Packet Loss}
We consider a system where transmission errors may result in the loss of messages. Given messages numbered by $n \in \mathbb{N}$ we define the process $I(n)$ where $I(n)=1$ if message $n$ is received error-free and $I(n)=0$ if message $n$ is erroneous. Considering age of information, we assume that an error-free message $n$ replaces all previous, potentially erroneous messages $n-1, n-2, \dots \, .$ Hence, for fluid and erroneous input $x$, we can define the packetized, error-free output
\begin{equation*}
P_{LI}(x) = \max_{n \in \mathbb{N}_0} \bigl\{ L(n) 1_{\{L(n) \le x\}} I(n) \bigr\} ,
\end{equation*}
with cumulative packet length $L(n)$, and $L(0),I(0) = 0$. With this definition we can make use of some basic results for packetizers~\cite{leboudec:networkcalculus, chang:performanceguarantees}. If the sequence of errors (number of consecutive zeros in $I(n)$) has an upper limit $\eta$, we have
\begin{equation*}
x \ge P_{LI}(x) \ge P_L(x) - \eta l_{\max} \ge x - (\eta+1) l_{\max} ,
\end{equation*}
where $P_L$ is defined in~\eqref{eq:packetizer}. Note that $P_L(P_{LI}(x)) = P_{LI}(x)$, but not the other way around.

Next, we consider a fluid system with service process $S(\tau,t)$ and packetized arrivals $A(t) = P_L(A(t))$. The system may introduce errors into the departure process $D(t) \ge A \otimes S(t)$. Subsequently, $D(t)$ is fed into a packetizer $P_{LI}$ with error detection. For the packetized error-free departure process we know that
\begin{equation*}
P_{LI}(D(t)) \ge P_{LI}(A \otimes S (t)) .
\end{equation*}
Since $P_{LI}(x)$ is non-decreasing and right-continuous
\begin{align*}
P_{LI}(D(t)) & \ge \inf_{\tau \in [0,t]} \bigl\{ P_{LI}(A(\tau) + S(\tau,t)) \bigr\} \\
& \ge \inf_{\tau \in [0,t]} \bigl\{ P_{L}(A(\tau) + S(\tau,t)) \bigr\} - \eta l_{\max}.
\end{align*}
Finally, with $P_L(x) \ge x - l_{\max}$ we obtain
\begin{align}
P_{LI}(D(t)) & \ge \inf_{\tau \in [0,t]} \bigl\{ A(\tau) + [S(\tau,t) - l_{\max}]_+ \bigr\} - \eta l_{\max} \nonumber \\
& = A \otimes [S - l_{\max}]_+ (t) - \eta l_{\max},
\label{eq:servicepacketizererror}
\end{align}
where we used that $P_{L}(A(\tau) + S(\tau,t)) \ge P_{L}(A(\tau)) = A(\tau)$ since $S(\tau,t)$ is non-negative and $A(t)$ is $P_L$ packetized to add the non-negativity condition $[.]_+$ that leads to service process $S_{P_L}(\tau,t) = [S(\tau,t) - l_{\max}]_+$ as in~\eqref{eq:servicepacketizer} before.

By insertion of~\eqref{eq:servicepacketizererror} into the definition of age of information~\eqref{eq:aoi_def} and following the same steps as in Sec.~\ref{sec:worstcase} we have
\begin{multline}
\Delta_{\max} \le \sup \biggl\{ \delta \ge 0 :  \inf \Bigl \{ \inf_{\tau \ge \delta} \{ \mathcal{S}_{P_L}(\tau) - \overline{\mathcal{E}}(\tau-\delta) \} , \\
\inf_{\tau \in [0,\delta)} \{ \mathcal{S}_{P_L}(\tau) + \underline{\mathcal{E}}(\delta-\tau) \} \Bigr \} \le \eta l_{\max} \biggr\} .
\label{eq:aoi_det_loss}
\end{multline}
For periodic update messages and a rate server with packetizer, see Sec.~\ref{sec:periodicupdates}, we have for $c \ge l/w$ for stability that
\begin{equation*}
\Delta_{\max} \le (\eta + 1) w + \frac{l}{c} .
\end{equation*}
Similarly, if the number of consecutive erroneous messages has a statistical upper bound $\eta_{\varepsilon}$ that is exceeded at most with probability $\varepsilon$, then $\Delta_{\varepsilon} = (\eta_{\varepsilon} +1)w +l/c$ is a statistical bound that satisfies $\mathsf{P} [\Delta(t) > \Delta_{\varepsilon}] \le \varepsilon$.
%
%
\section{Random Service}
\label{sec:randomservice}
We consider systems with a random service, such as a wireless channel, where the state of the channel determines the success of a transmission, or a scheduler with cross-traffic.
%
%
\subsection{Statistical Service Curve}
To model systems with a random service, we employ basic results from the stochastic network calculus as outlined in~\cite{fidler:netcalcguide}. It is common in stochastic network calculus~\cite{jiang:stochasticnetworkcalculus,fidler:netcalcguide} to use a discrete time model. Continuous time can be converted to discrete time by sampling~\cite{leboudec:networkcalculus}. We will use the technique from~\cite{ciucu:networkservicecurvescaling2} to estimate the sampling error.

A stochastic extension of the service envelope~\eqref{eq:servicecurve} is
\begin{equation}
\mathsf{P}[\exists \tau \in [0,t]: S(\tau,t) < \mathcal{S}_{\varepsilon}(t-\tau) ] \le \varepsilon ,
\label{eq:serviceenvelope}
\end{equation}
for all $t \ge 0$. The envelope function $\mathcal{S}_{\varepsilon}(t) \in \mathcal{F}_0$ provides a statistical service guarantee that has probability of underflow $\varepsilon \in [0,1]$, where $\varepsilon$ is typically small, e.g., $10^{-6}$. It is important to note that~\eqref{eq:serviceenvelope} considers the probability of underflow along an entire sample path $\tau \in [0,t]$. Thus, it can be directly inserted into~\eqref{eq:serviceprocess}. It follows that \mbox{$\mathsf{P}[D(t) < A \otimes \mathcal{S}_{\varepsilon}(t)] \le \varepsilon$}, where $\mathcal{S}_{\varepsilon}(t)$ (with a certain parameterization) is known as statistical service curve~\cite{ciucu:networkservicecurvescaling2} or weak stochastic service curve~\cite{jiang:stochasticnetworkcalculus}. With~\eqref{eq:aoi} we obtain a statistical age of information bound
\begin{equation*}
\mathsf{P} [\Delta(t) > \Delta_{\varepsilon}] \le \varepsilon ,
\end{equation*}
for $t \ge 0$ where $\Delta_{\varepsilon}$ is given by substitution of $\mathcal{S}_{\varepsilon}(\tau)$ into~\eqref{eq:aoi_det}.

To derive $\varepsilon$ for a certain function $\mathcal{S}_{\varepsilon}(t)$, we use a bound of the moment generating function of $S(\tau,t)$ defined as
\begin{equation}
\mathsf{E} \bigl[e^{-\theta S(\tau,t)}\bigr] \le e^{-\theta(\rho(\theta) (t-\tau) - \sigma(\theta))} ,
\label{eq:envelopeparameters}
\end{equation}
for $\theta \ge 0$ and envelope parameters $\sigma(\theta) \ge 0$ and $\rho(\theta) > 0$. The bound is a variant of the $(\sigma(\theta),\rho(\theta))$ traffic characterization of~\cite{chang:performanceguarantees} applied to service processes, see~\cite{fidler:netcalcguide} for details.

We define a sampling interval $\tau_0 > 0$ and number the intervals by $\kappa \in \mathbb{N}_0$, i.e., the interval $[0,t]$ is extended to the left and divided into subintervals $[t-(\kappa+1)\tau_0,t-\kappa\tau_0]$ for $\kappa \in [0,\lceil t/\tau_0\rceil - 1]$. Since $\mathcal{S}_{\varepsilon}(t) \in \mathcal{F}_0$ is non-decreasing and $S(\tau,t)$ is non-negative and non-increasing with increasing $\tau$, we have for $\kappa \in [0,\lceil t/\tau_0 \rceil-1]$ that if
\begin{align*}
& S(t-\kappa\tau_0,t) \ge \mathcal{S}_{\varepsilon}((\kappa+1)\tau_0) \\
\Rightarrow & S(\tau,t) \ge \mathcal{S}_{\varepsilon}(t-\tau), \, \forall \tau \in [t-(\kappa+1)\tau_0,t-\kappa\tau_0] .
\end{align*}
Hence, with application of the union bound
\begin{align}
& \mathsf{P}[\exists \tau \in [0,t]: S(\tau,t) < \mathcal{S}_{\varepsilon}(t-\tau) ] \nonumber \\
\le & \sum_{\kappa=0}^{\lceil t/\tau_0 \rceil - 1} \mathsf{P} [S(t-\kappa\tau_0,t) < \mathcal{S}_{\varepsilon}((\kappa+1)\tau_0) ] .
\label{eq:sumofintervals}
\end{align}
We choose the shape of the service curve
\begin{equation*}
\mathcal{S}_{\varepsilon}(t) = [r t - b]_+
\end{equation*}
with $0 < r < \rho(\theta)$ and $b \ge r \tau_0 + \sigma(\theta)$. It follows that $\mathcal{S}_{\varepsilon}(\tau_0) = 0$ and the summand for $\kappa=0$ in~\eqref{eq:sumofintervals} is zero trivially since $S(\tau,t)$ is non-negative. For the same reason, we can drop the $[.]_+$ condition of $\mathcal{S}_{\varepsilon}(t)$ in the following step.

To estimate~\eqref{eq:sumofintervals}, we make use of Chernoff's lower bound $\mathsf{P}[X \le x] \le e^{\theta x} \mathsf{E}[e^{-\theta X}]$ for a random variable $X$ and $\theta > 0$. We insert $S(t-\kappa\tau_0,t)$ for $X$, $\mathcal{S}_{\varepsilon}((\kappa+1)\tau_0)$ for $x$, and  estimate $\mathsf{E}[e^{-\theta S(t-\kappa\tau_0,t)}]$ by~\eqref{eq:envelopeparameters} to obtain
\begin{align}
& \mathsf{P}[\exists \tau \in [0,t]: S(\tau,t) < \mathcal{S}_{\varepsilon}(t-\tau) ] \nonumber \\
\le & \sum_{\kappa=1}^{\lceil t/\tau_0 \rceil - 1} e^{-\theta (\rho(\theta)\kappa\tau_0 - \sigma(\theta))} e^{\theta (r (\kappa+1) \tau_0 - b)}  \nonumber \\
\le & e^{-\theta(b-r\tau_0-\sigma(\theta))} \sum_{\kappa=1}^{\infty} e^{-\theta (\rho(\theta)-r)\kappa\tau_0} \nonumber \\
\le & e^{-\theta(b-r\tau_0-\sigma(\theta))} \int_{0}^{\infty} e^{-\theta (\rho(\theta) - r)x\tau_0} dx \nonumber \\
= & \frac{e^{-\theta (b - r \tau_0-\sigma(\theta))}}{\theta (\rho(\theta)-r)\tau_0} .
\label{eq:epsilon}
\end{align}
In the third line, we let $t \rightarrow \infty$, and in the fourth line we made use of the fact that the summands are decreasing with $\kappa$ since $\rho(\theta) > r > 0$, so that each summand is bounded from above by an integral of unit size left of that summand. Equating~\eqref{eq:epsilon} with $\varepsilon$ we can solve for
\begin{equation}
b = - \frac{1}{\theta} \ln ( \theta (\rho(\theta) - r)\tau_0\varepsilon ) + r\tau_0 + \sigma(\theta).
\label{eq:underflowprofile}
\end{equation}

Below, we insert the envelope parameters $(\sigma(\theta),\rho(\theta))$ of the service process. The parameters of a variety of service models are known, including wireless channels~\cite{fidler:radiocalculus} and scheduling algorithms~\cite{li:effectivebandwidthcalculus2}, which we will employ in Sec.~\ref{sec:markovchannel} and Sec.~\ref{sec:scheduling}, respectively.

Lastly, we include packetization. Note that if $D(t) \ge A \otimes \mathcal{S}_{\varepsilon}(t)$ then also $P_L(D(t)) \ge P_L( A \otimes \mathcal{S}_{\varepsilon}(t))$ and following the steps of the derivation of~\eqref{eq:servicecurvepacketizer}, $\mathcal{S}_{\varepsilon,P_L}(t) =  [[rt - b]_+ -l_{\max}]_+$ is a statistical service curve of the series of the system with random service and the packetizer. As before, we express this as a latency-rate server
\begin{equation*}
\mathcal{S}_{\varepsilon,P_L}(t) = r [t - (b+l_{\max})/r]_+
\end{equation*}
with latency $t_0 = (b+l_{\max})/r$.

Using the representation as a latency-rate curve, we can directly apply the solution~\eqref{eq:lraoi} for periodic updates defined in Sec.~\ref{sec:periodicupdates}, to find the statistical age of information bound
\begin{equation}
\Delta_{\varepsilon} = w + \frac{b+l}{r}
\label{eq:aoistochastic}
\end{equation}
under the stability condition $l/w \le r < \rho(\theta)$ and $\theta > 0$.
%
%
\subsection{Markov Channel}
\label{sec:markovchannel}
An established model of a wireless channel with memory is the Gilbert-Elliott model that uses a Markov chain to classify the state of the channel. Each state has a defined transmission rate that is determined by  the modulation and coding scheme. Here, we use a Markov on-off model with transition rates \mbox{$\mu > 0$} from on to off state and $\lambda > 0$ from off to on state. The probability of the on state is $p_{\text{on}} = \lambda/(\lambda+\mu)$ and $p_{\text{off}} = 1-p_{\text{on}}$. The memory of the channel is characterized by the burstiness parameter $\beta = 1/\lambda + 1/\mu$ that is the mean time to change state twice. The transmission rate in on state is $c$, and in off state $0$. The channel has mean rate $\gamma = c p_{\text{on}}$, and for $\theta > 0$ envelope parameters $\sigma(\theta) = 0$ and~\cite{gibbens:markovonoff,kelly:effectivebandwidths,fidler:radiocalculus}
\begin{equation}
\rho(\theta) = - \frac{1}{2\theta} \left(\sqrt{(\lambda - \mu -\theta c)^2 + 4\lambda\mu} -\lambda - \mu - \theta c \right) .
\label{eq:markovchannel}
\end{equation}

For a numerical evaluation we insert~\eqref{eq:markovchannel} into~\eqref{eq:underflowprofile} and then use~\eqref{eq:aoistochastic} for periodic updates. We optimize the free parameters $r \in [l/w, \rho(\theta))$, $\theta > 0$, and $\tau_0 > 0$ numerically to find the smallest statistical age of information bound $\Delta_{\varepsilon}$.

\begin{figure}
\centering
\includegraphics[width=0.95\linewidth]{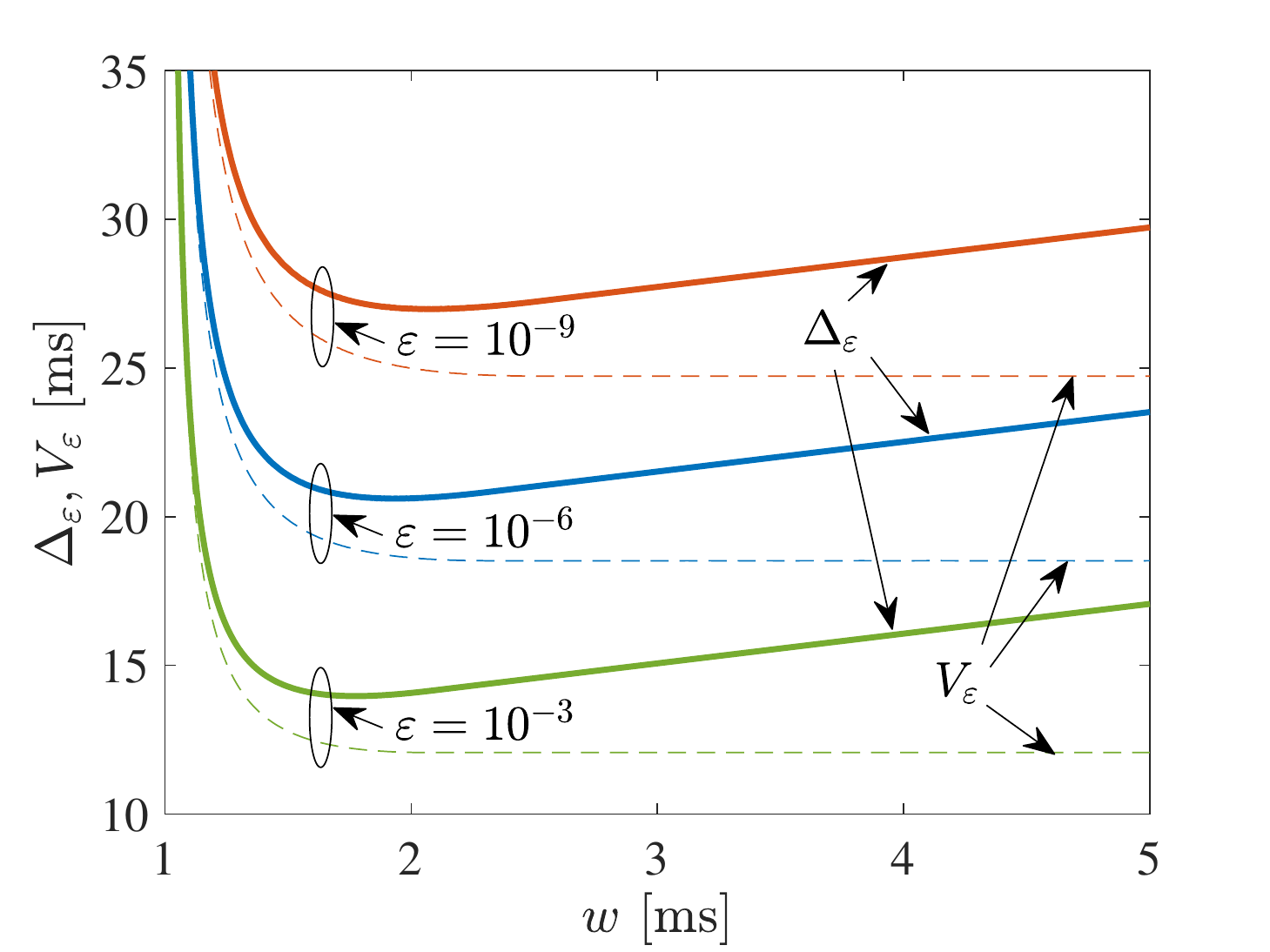}
\caption{Statistical age of information bound $\Delta_{\varepsilon}$ and virtual delay bound $V_{\varepsilon}$ with probability $\varepsilon$ for a Markov channel and message generation interval $w$.}
\label{fig:aoi_onoff_periodic}
\end{figure}
In Fig.~\ref{fig:aoi_onoff_periodic}, we show $\Delta_{\varepsilon}$~\eqref{eq:aoistochastic} for a periodic source with packet size $l = 1$~kb  and different update intervals $w$ in ms. The source is transmitted via a Markov on-off channel. The probability of the on state is $p_{\text{on}} = 0.9$, the burstiness of the channel is $\beta = 8$~ms, and the mean rate of the channel is $\gamma = 1$~Mb/s. We use $\varepsilon \in \{10^{-3}, 10^{-6}, 10^{-9}\}$. For comparison, we also include the statistical virtual delay bound $V_{\varepsilon}$.

Regarding~\eqref{eq:aoistochastic}, $\Delta_{\varepsilon}$ depends in two ways on $w$. Firstly, $\Delta_{\varepsilon}$ grows linearly with $w$. This is clearly visible in Fig.~\ref{fig:aoi_onoff_periodic} where $\Delta_{\varepsilon}$ but not $V_{\varepsilon}$ increases if $w$ becomes large. Secondly, $\Delta_{\varepsilon}$ and $V_{\varepsilon}$ are affected by delays that occur during off periods of the channel. This is expressed by parameter $b$. The effect that $w$ has on $b$ and thus on $\Delta_{\varepsilon}$ and $V_{\varepsilon}$ is via the stability condition $l/w \le r < \rho(\theta)$, where small $w$ implies small $\theta$ resulting in large $b$~\eqref{eq:underflowprofile}. This corresponds to queueing that arises during off periods of the channel if the utilization is high. This effect is visible in Fig.~\ref{fig:aoi_onoff_periodic} for $\Delta_{\varepsilon}$ and $V_{\varepsilon}$ if $w$ becomes small.

\begin{figure}
\subfigure[$w=2$]{
\includegraphics[width=0.46\linewidth]{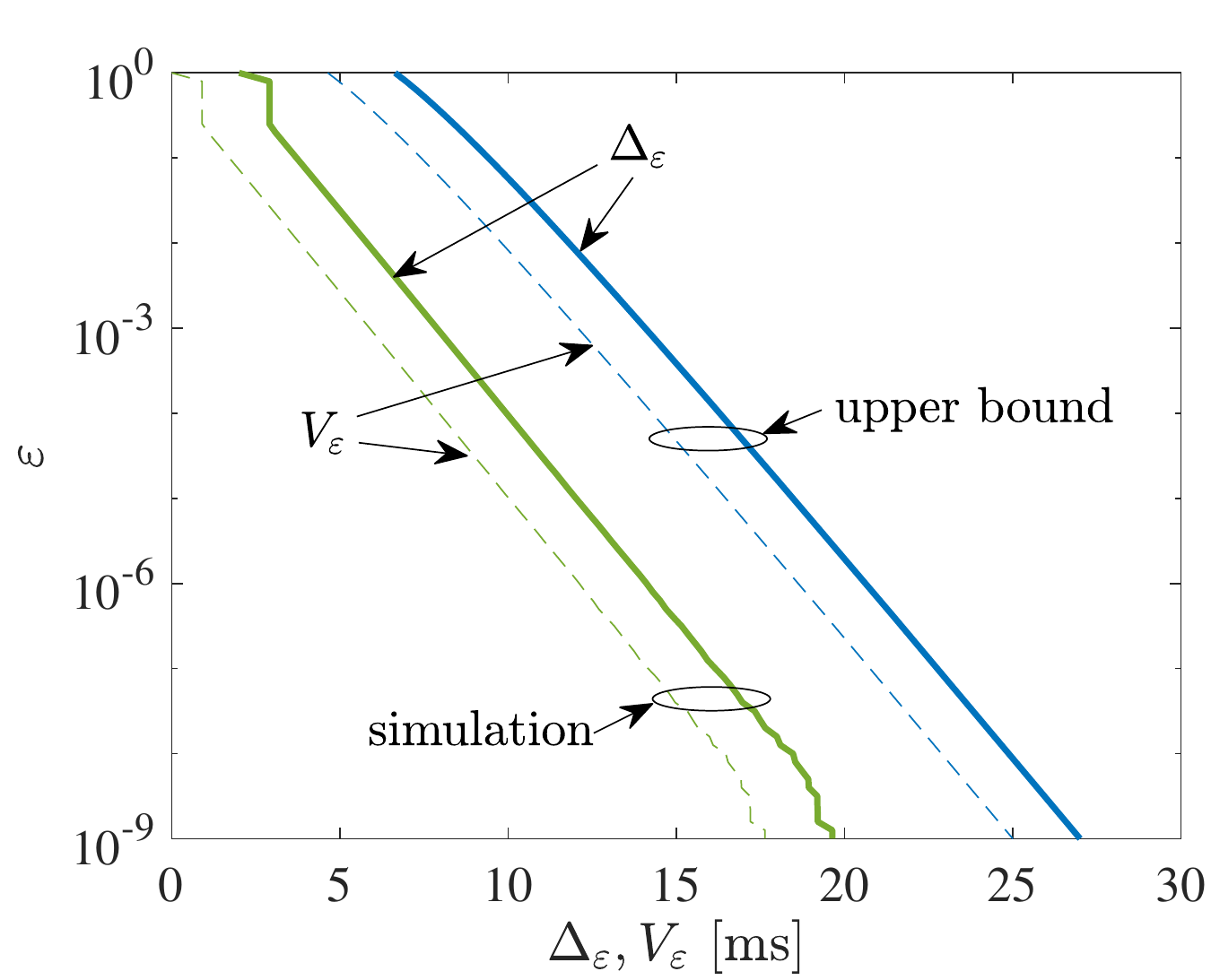}
\label{fig:aoi_onoff_periodic_sim_eps}
}
\hfill
\subfigure[$\varepsilon=10^{-6}$]{
\includegraphics[width=0.46\linewidth]{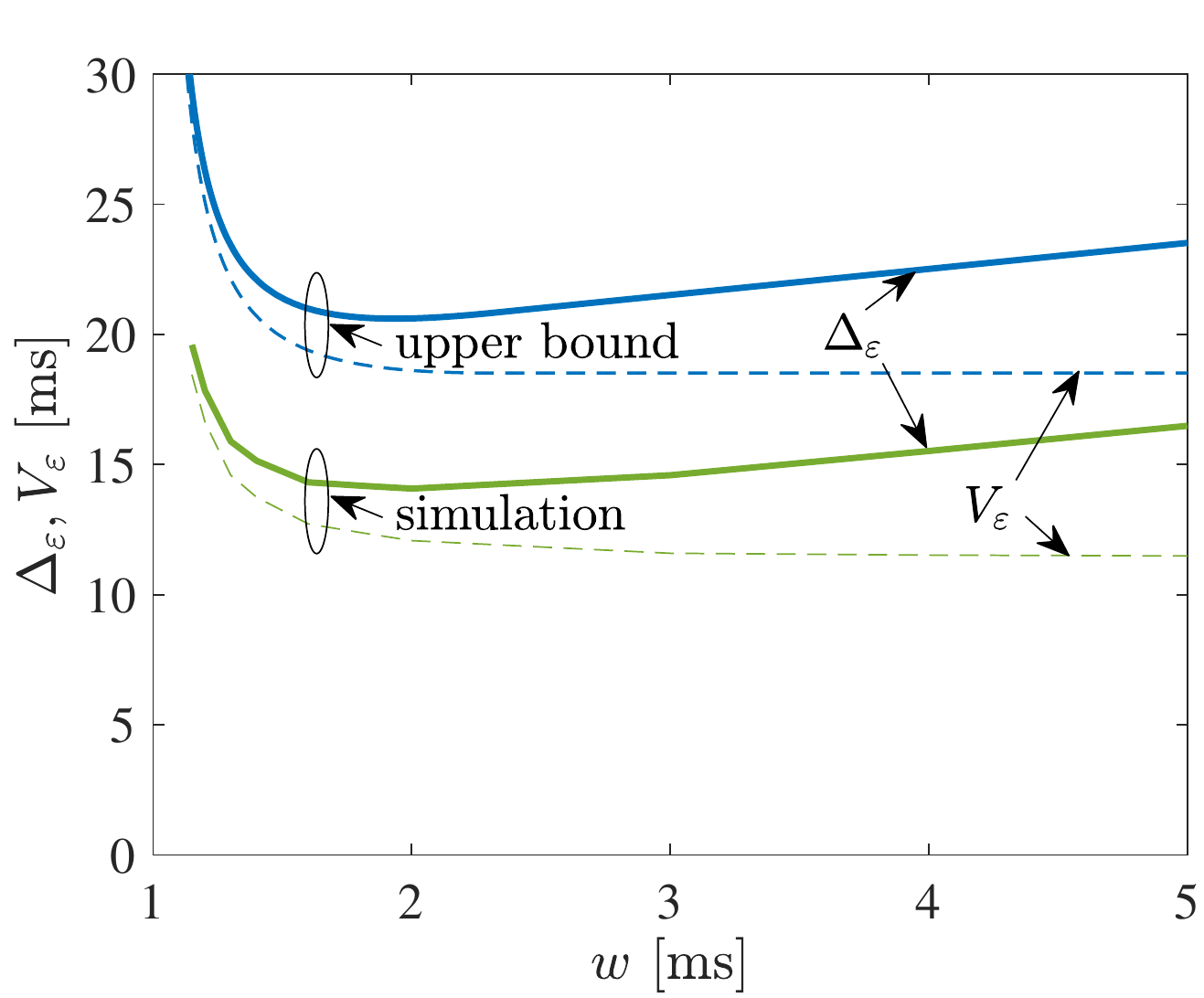}
\label{fig:aoi_onoff_periodic_sim_w}
}
\caption{Statistical age of information bound $\Delta_{\varepsilon}$ and virtual delay bound $V_{\varepsilon}$ for the same system as in Fig.~\ref{fig:aoi_onoff_periodic} compared to simulation results.}
\label{fig:aoi_onoff_periodic_sim}
\end{figure}
Fig.~\ref{fig:aoi_onoff_periodic_sim} compares simulation results with the statistical bounds depicted in Fig~\ref{fig:aoi_onoff_periodic}. In the simulations, we generated $10^9$ samples of age of information and delay for each choice of the parameter $w$. In Fig.~\ref{fig:aoi_onoff_periodic_sim_eps} we show the tail decay for $w=2$ and in Fig.~\ref{fig:aoi_onoff_periodic_sim_w} we plot the respective $1-\varepsilon$-quantiles for $\varepsilon = 10^{-6}$ and varying $w$. Clearly, the analysis gives upper bounds that match the speed of tail decay with $\varepsilon$ of the simulation results and reproduce the correct trend with $w$.
%
%
\section{Random Arrivals}
\label{sec:randomarrivals}
So far, we considered periodic updates that are common in time-triggered systems. Event-triggered systems, on the other hand, generate update messages in case a defined event occurs, e.g., if a sensor reading exceeds a certain threshold. The occurrence of events can be modeled as a random process.

The analysis of random updates is dual to that of random service in Sec.~\ref{sec:randomservice}. We reuse some of the notation, where $\varepsilon_A$, $\rho_A$, $\sigma_A$, $r_A$, and $b_A$ refer to the arrivals $A(t)$. For readability, we skip the subscript $A$ where it is not ambiguous.
%
%
\subsection{Statistical Upper Arrival Envelope}
A statistical version of the upper arrival envelope~\eqref{eq:envelope} is
\begin{equation*}
\mathsf{P}[\exists \tau \in [0,t]: A(\tau,t) > \overline{\mathcal{E}}_{\varepsilon}(t-\tau)] \le \overline{\varepsilon} ,
\end{equation*}
for all $t \ge 0$ and $\overline{\mathcal{E}}_{\varepsilon}(t) \in \mathcal{F}_0$. The definition of envelope considers the probability of overflow $\overline{\varepsilon} \in [0,1]$ along an entire sample path, as discussed previously for~\eqref{eq:serviceenvelope}.

In the following derivation, we relax the requirement that $\overline{\mathcal{E}}_{\varepsilon}(t) \in \mathcal{F}_0$ and permit $\overline{\mathcal{E}}_{\varepsilon}(0) > 0$ for notational simplicity. We note that $\overline{\mathcal{E}}_{\varepsilon}(0) = 0 \ge A(t,t)$ holds trivially since $A(t,t) = 0$ for all $t$.

We use the $(\sigma(\theta),\rho(\theta))$ traffic characterization of~\cite{chang:performanceguarantees}, i.e.,
\begin{equation*}
\mathsf{E} \bigl[e^{\theta A(\tau,t)}\bigr] \le e^{\theta(\rho(\theta) (t-\tau) + \sigma(\theta))} ,
\end{equation*}
for $\theta \ge 0$, $\sigma(\theta) \ge 0$, and $\rho(\theta) > 0$, that is dual to~\eqref{eq:envelopeparameters}.

Employing a discretization step as in Sec.~\ref{sec:randomservice}, we have for $A(t)$ and $\overline{\mathcal{E}}_{\varepsilon}(t)$ non-decreasing, $\tau_0 > 0$, and $\kappa \in \mathbb{N}$, where $\kappa \in [1,\lceil t/\tau_0 \rceil]$ that if
\begin{align*}
& A(t-\kappa\tau_0,t) \le \overline{\mathcal{E}}_{\varepsilon}((\kappa-1)\tau_0) \\
\Rightarrow & A(\tau,t) \le \overline{\mathcal{E}}_{\varepsilon}(t-\tau), \, \forall \tau \in [t-\kappa\tau_0,t-(\kappa-1)\tau_0] .
\end{align*}
With the union bound we have
\begin{align*}
& \mathsf{P}[\exists \tau \in [0,t]: A(\tau,t) > \overline{\mathcal{E}}_{\varepsilon}(t-\tau) ] \\
\le & \sum_{\kappa=1}^{\lceil t/\tau_0 \rceil} \mathsf{P} [A(t-\kappa\tau_0,t) > \overline{\mathcal{E}}_{\varepsilon}((\kappa-1)\tau_0) ] .
\end{align*}
We choose
\begin{equation*}
\overline{\mathcal{E}}_{\varepsilon}(t) = r t + b ,
\end{equation*}
where $0 < \rho(\theta) < r$ and $b \ge r \tau_0 + \sigma(\theta)$. By use of Chernoff's upper bound $\mathsf{P}[X \ge x] \le e^{-\theta x} \mathsf{E}[e^{\theta X}]$ for a random variable $X$ and $\theta > 0$ it follows that
\begin{align}
& \mathsf{P}[\exists \tau \in [0,t]: A(\tau,t) > \overline{\mathcal{E}}_{\varepsilon}(t-\tau) ] \nonumber \\
\le & \sum_{\kappa=1}^{\lceil t/\tau_0 \rceil} e^{\theta (\rho(\theta)\kappa\tau_0 + \sigma(\theta))} e^{-\theta (r (\kappa-1) \tau_0 + b)}  \nonumber \\
\le & e^{-\theta(b-r\tau_0-\sigma(\theta))} \int_{0}^{\infty} e^{-\theta (r-\rho(\theta) )x\tau_0} dx \nonumber \\
= & \frac{e^{-\theta (b - r \tau_0-\sigma(\theta))}}{\theta (r-\rho(\theta))\tau_0} .
\label{eq:epsilonarrival}
\end{align}
We equate~\eqref{eq:epsilonarrival} with $\overline{\varepsilon}$, solve for
\begin{equation}
b = - \frac{1}{\theta} \ln (\theta (r-\rho(\theta)) \tau_0\overline{\varepsilon}) + r\tau_0 + \sigma(\theta),
\label{eq:overflowprofile}
\end{equation}
and substitute $\overline{\mathcal{E}}_{\varepsilon}(t) = rt + b$ into the age of information~\eqref{eq:aoi_det}.
%
%
\subsection{Statistical Lower Arrival Envelope}
Further,~\eqref{eq:aoi_det} uses a lower arrival envelope. A statistical lower envelope $\underline{\mathcal{E}}_{\varepsilon}(t) \in \mathcal{F}_0$ that fits in with~\eqref{eq:aoi} has the form
\begin{equation}
\mathsf{P}[\exists t \ge \tau: A(\tau,t) < \underline{\mathcal{E}}_{\varepsilon}(t-\tau)] \le \underline{\varepsilon},
\label{eq:lowerarrivalenvelope}
\end{equation}
for $\tau \ge 0$. Instead of using the approach as above to derive an envelope function, we observe that the age of information~\eqref{eq:aoi_det} depends only on the first non-zero value of $\underline{\mathcal{E}}_{\varepsilon}(t)$. Hence, we define
\begin{equation*}
\underline{\mathcal{E}}_{\varepsilon}(t) = 1_{\{t>u\}} l_{\min}
\end{equation*}
where $l_{\min} > 0$ is the minimal packet size and parameter $u>0$ is a statistical measure of the time until the next update occurs\footnote{In case of a system with packet loss,~\eqref{eq:aoi_det_loss} considers the first value that exceeds $\eta l_{\max}$ and the envelope $\underline{\mathcal{E}}_{\varepsilon}(t) = 1_{\{t>u\}}(\l_{\min} + \eta l_{\max})$ is used instead.}.
By insertion into~\eqref{eq:lowerarrivalenvelope} it follows for $A(t)$ non-decreasing that
\begin{equation*}
\mathsf{P}[\exists t \ge \tau: A(\tau,t) < 1_{\{t-\tau>u\}} l_{\min}] \le \mathsf{P}[A(\tau,\tau+u) < l_{\min}] .
\end{equation*}
We equate the right hand side with $\underline{\varepsilon}$ and apply the desired probability distribution, e.g., Poisson, in Sec.~\ref{sec:poisson}.

To obtain the age of information, we insert \mbox{$\overline{\mathcal{E}}_{\varepsilon}(t) = rt + b$}, and $\underline{\mathcal{E}}_{\varepsilon}(t) = 1_{\{t>u\}} l_{\min}$, as well as the deterministic or statistical latency rate server model $\mathcal{S}(t) = c[t-t_0]_+$ into~\eqref{eq:aoi_det}
\begin{multline*}
\Delta_{\varepsilon} \le \sup \biggl\{ \delta \ge 0 :  \inf \Bigl \{ \inf_{\tau \ge \delta} \{ c[\tau-t_0]_+ - r (\tau-\delta) - b \} , \\
\inf_{\tau \in [0,\delta)} \{ c[\tau-t_0]_+ + 1_{\{\delta-\tau>u\}} l_{\min} \} \Bigr \} \le 0 \biggr\} ,
\end{multline*}
where $\varepsilon = \overline{\varepsilon} + \underline{\varepsilon}$ by the union bound. With $c \ge r$ for stability, it follows that
\begin{equation}
\Delta_{\varepsilon} \le \max \biggl\{ \frac{b}{c}, u \biggr\} + t_0 .
\label{eq:aoistochasticarrivals}
\end{equation}
%
%
\subsection{Poisson Arrivals}
\label{sec:poisson}
To evaluate~\eqref{eq:aoistochasticarrivals} we use arrivals with constant packet length $l > 0$ and exponential inter-arrival times with mean value $w > 0$. Hence, the arrivals form a Poisson process with arrival rate $1/w$. The envelope parameters of the Poisson process are $\sigma(\theta)=0$ and
\begin{equation}
\rho(\theta) = \frac{e^{\theta l} - 1}{\theta w} ,
\label{eq:rhopoisson}
\end{equation}
for $\theta > 0$~\cite{fidler:netcalcguide}. By insertion of $\rho(\theta) < r$ into~\eqref{eq:overflowprofile} and choice of $\overline{\varepsilon}$, parameter $b$ is determined and can be inserted into~\eqref{eq:aoistochasticarrivals}. To fix the remaining parameter $u$, we use the Poisson distribution
\begin{equation*}
\mathsf{P}[A(\tau,\tau+t) = \eta l] = \frac{e^{-\frac{t}{w}} (\frac{t}{w})^\eta}{\eta!} ,
\end{equation*}
for $\eta \ge 0$. With $\underline{\varepsilon} = \mathsf{P}[A(\tau,\tau+u) < l] = \mathsf{P}[A(\tau,\tau+u) = 0]$ $= e^{-u/w}$ we can solve for $u$ and insert $u = -w \ln \underline{\varepsilon}$ into~\eqref{eq:aoistochasticarrivals}.

\begin{figure}
\centering
\includegraphics[width=0.93\linewidth]{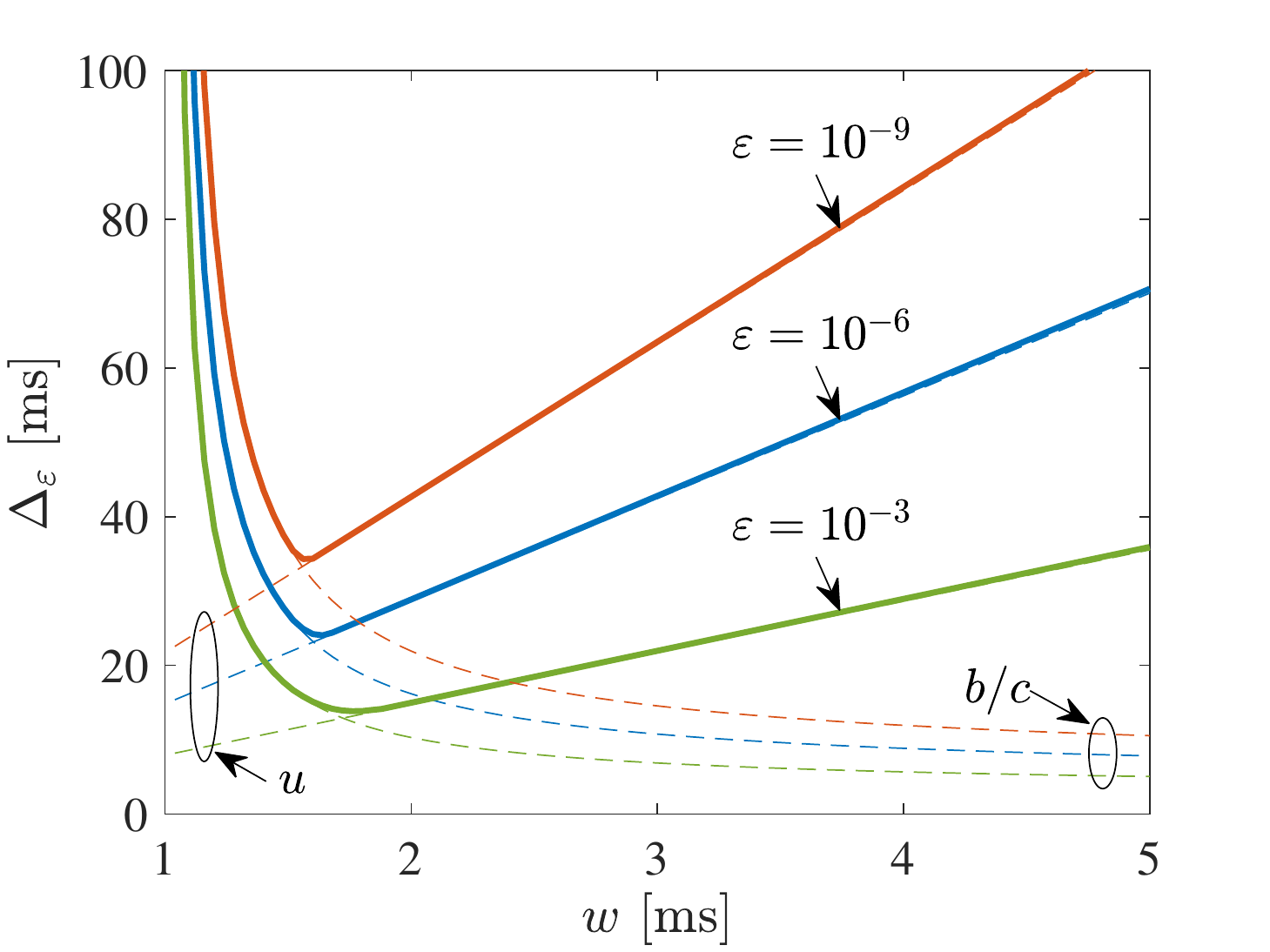}
\caption{Statistical age of information bound $\Delta_{\varepsilon}$ for Poisson arrivals. For small $w$ the age of information is mostly due to congestive queueing in the network, whereas for large $w$ it is due to idle waiting.}
\label{fig:aoi_lr_poisson}
\end{figure}

In Fig.~\ref{fig:aoi_lr_poisson} we depict the age of information~\eqref{eq:aoistochasticarrivals} for a system with capacity $c=1$~Mb/s, packet length $l=1$~kb, $t_0 = l/c = 1$~ms and $\varepsilon \in \{10^{-3}, 10^{-6}, 10^{-9}\}$. We optimize parameters $r \in (\rho(\theta),c]$, $\theta > 0$, and $\tau_0 > 0$ numerically and select $\overline{\varepsilon}, \underline{\varepsilon} \in [0,1]$ so that $\overline{\varepsilon} + \underline{\varepsilon} = \varepsilon$ and $b/c = u$.

Fig.~\ref{fig:aoi_lr_poisson} clearly shows two regimes that govern the age of information. For small $w$, congestive queueing delays are dominant that are derived from the upper arrival envelope in the first line of~\eqref{eq:aoi_det} corresponding to the term $b/c$ in~\eqref{eq:aoistochasticarrivals}. For large $w$, the age of information is mostly due to idle waiting represented by the lower arrival envelope in the second line of~\eqref{eq:aoi_det} and parameter $u$ in~\eqref{eq:aoistochasticarrivals}, respectively. The optimal update interval $w$ strikes a balance between these two effects.
%
%
\section{Multiplexing and Scheduling}
\label{sec:scheduling}
The min-plus representation of the network calculus naturally extends to tandem systems, including multiplexing and scheduling of traffic flows. Here, we consider several sources with different priorities that send periodic updates via a wireless channel.
\subsection{Priority Scheduler}
We consider a work-conserving system with service process $S(\tau,t)$ and service parameters $(\sigma_S(\theta),\rho_S(\theta))$. The system serves a number of independent and identically distributed (iid) flows $A_i(t)$ for $i \in \mathbb{N}$ that are numbered in descending order of priority. We analyze the age of information of flow $m+1$, i.e., there are $m$ flows of higher priority. Given each of the flows has the envelope parameters $(\sigma_A(\theta),\rho_A(\theta))$, it is known that the aggregate of $m$ flows has parameters $(m\sigma_A(\theta),m\rho_A(\theta))$. Further, the service that is left over for flow $m+1$ satisfies the definition of service process~\eqref{eq:serviceprocess} with envelope parameters $(\sigma_S(\theta)+m\sigma_A(\theta),\rho_S(\theta)-m\rho_A(\theta))$~\cite{fidler:netcalcguide}. The parameters are inserted into~\eqref{eq:underflowprofile} to derive the age of information as in Sec.~\ref{sec:randomservice}. We will use the Markov channel model from Sec.~\ref{sec:markovchannel} with $\sigma_S(\theta) = 0$ and $\rho_S(\theta)$ given by~\eqref{eq:markovchannel}. For random updates, it is straightforward to insert $\sigma_A(\theta)=0$ and $\rho_A(\theta)$ given by~\eqref{eq:rhopoisson} for the Poisson process. Periodic updates will be considered in the next subsection.
\subsection{Periodic Updates}
For the moment generating function of a periodic source with packet length $l>0$ and update interval width $w>0$ it is known that~\cite{kelly:effectivebandwidths}
\begin{equation*}
\mathsf{E}\bigl[e^{\theta A(\tau,\tau+t)}\bigr] = \exp \biggl( \theta l \Bigl\lfloor \frac{t}{w} \Bigr\rfloor + \ln \biggl(1 + \Bigl(\frac{t}{w} - \Bigl\lfloor \frac{t}{w} \Bigr\rfloor \Bigr)(e^{\theta l}-1)\biggr)\!\biggr),
\end{equation*}
for $\theta > 0$. The moment generating function considers sources with statistically independent, uniformly distributed phase. We choose parameter $\rho_A = l/w$, that is the mean rate, and compute $\sigma_A(\theta) = \sup_{\tau \in [0,t]} \{\ln (\mathsf{E}[e^{\theta A(\tau,\tau+t)}])/\theta -\rho_A t \}$.

We consider iid periodic sources numbered in descending order of priority that are transmitted via a Markov channel. The age of information of source $m+1$ is given by~\eqref{eq:aoistochastic}, where we insert~\eqref{eq:underflowprofile} with parameters $(m\sigma_A(\theta),\rho_S(\theta)-m\rho_A)$ and $\rho_S(\theta)$ of the Markov channel is given in~\eqref{eq:markovchannel}. The stability condition for source $m+1$ is $l/w \le r < \rho_S(\theta)-m\rho_A$.

\begin{figure}
\centering
\includegraphics[width=0.93\linewidth]{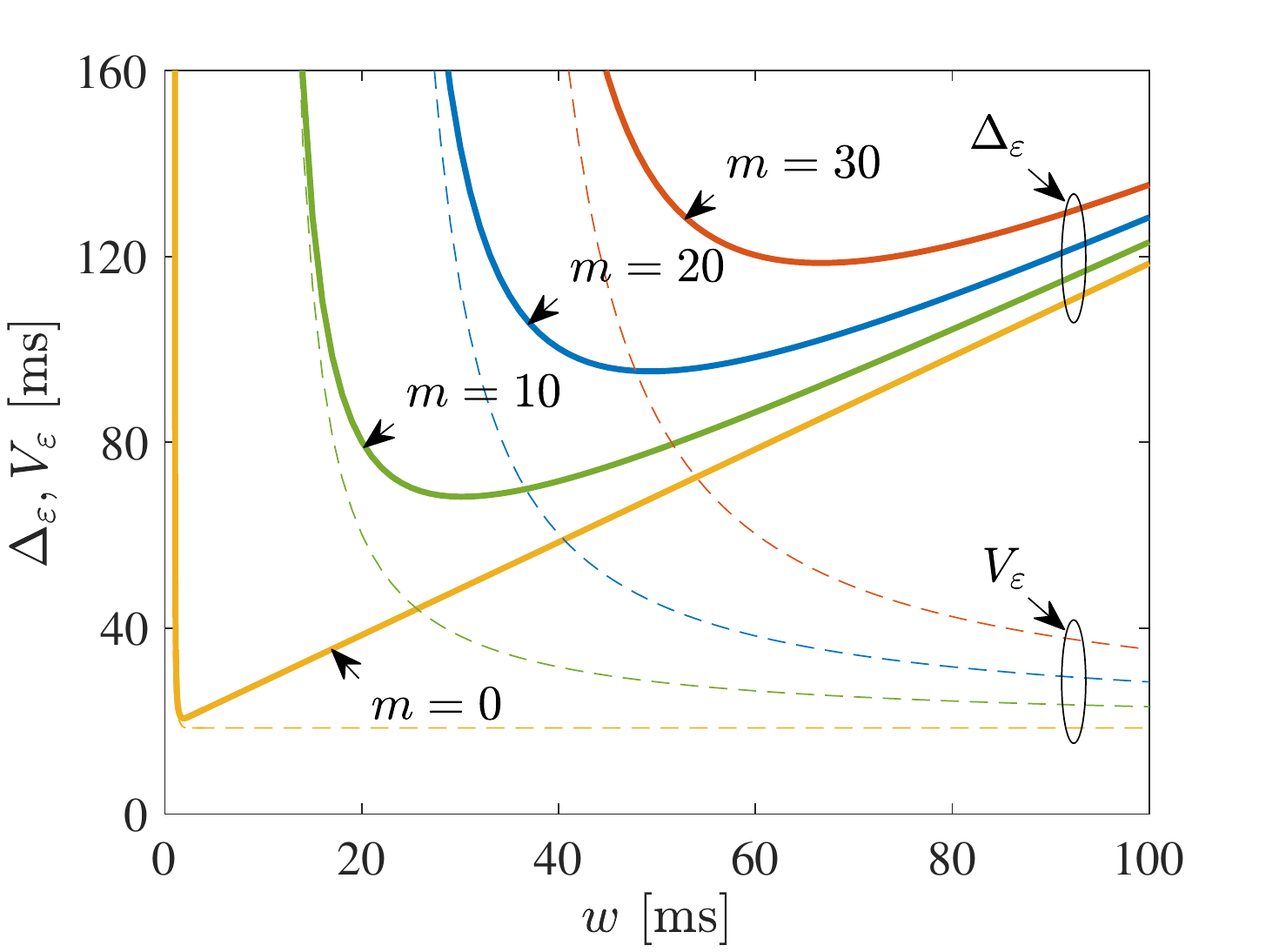}
\caption{Statistical age of information bound $\Delta_{\varepsilon}$ and virtual delay bound $V_{\varepsilon}$ for priority $m+1$.}
\label{fig:aoi_scheduling_onoff_periodic}
\end{figure}

In Fig.~\ref{fig:aoi_scheduling_onoff_periodic}, we show the age of information bound of source $m+1$ for $m \in \{0,10,20,30\}$. For the Markov channel we use the same parameters as for Fig.~\ref{fig:aoi_onoff_periodic}, i.e., $p_{\text{on}} = 0.9$, mean rate $\gamma = 1$~Mb/s, burstiness $\beta = 8$ and packet length $l=1$~kb. For comparison, we also include the corresponding bound of the virtual delay, where $\varepsilon = 10^{-6}$ for all curves.

The curve for $m=0$, i.e., for source $1$ is identical to the one in Fig.~\ref{fig:aoi_onoff_periodic}. Clearly, for source $1$, that has highest priority, a small update interval $w \approx 2$ minimizes the age of information. This leaves, however, little residual service, so that the age of information of lower priority sources does not converge. Larger update intervals $w$, on the other hand, achieve stability also for lower priorities, but increase the age of information due to less frequent updates. Interestingly, all curves converge with increasing $w$, i.e., in a moderately loaded system the priority of the sources has little impact on the age of information. This effect is due to statistical multiplexing gains that are put into effect since the phases of the sources are statistically independent.
%
%
\section{Conclusions}
\label{sec:conclusions}
We phrased age of information in the min-plus network calculus and derived worst-case and statistical age of information bounds. We obtained solutions for sources with periodic updates and random updates, respectively, and systems with a random service, including wireless channels and schedulers with random cross-traffic. We showed different effects how the update interval affects the age of information. Our results enable finding the optimal update interval that achieves the minimal age of information. Owing to the properties of the network calculus, our analysis can be easily extended to include further traffic and service models as well as multi-hop networks.
%
%
\balance
\bibliographystyle{IEEEtran}
\bibliography{IEEEabrv,IEEEfidler}

\begin{thebibliography}{10}
\providecommand{\url}[1]{#1}
\csname url@samestyle\endcsname
\providecommand{\newblock}{\relax}
\providecommand{\bibinfo}[2]{#2}
\providecommand{\BIBentrySTDinterwordspacing}{\spaceskip=0pt\relax}
\providecommand{\BIBentryALTinterwordstretchfactor}{4}
\providecommand{\BIBentryALTinterwordspacing}{\spaceskip=\fontdimen2\font plus
\BIBentryALTinterwordstretchfactor\fontdimen3\font minus
  \fontdimen4\font\relax}
\providecommand{\BIBforeignlanguage}[2]{{%
\expandafter\ifx\csname l@#1\endcsname\relax
\typeout{** WARNING: IEEEtran.bst: No hyphenation pattern has been}%
\typeout{** loaded for the language `#1'. Using the pattern for}%
\typeout{** the default language instead.}%
\else
\language=\csname l@#1\endcsname
\fi
#2}}
\providecommand{\BIBdecl}{\relax}
\BIBdecl

\bibitem{kaul:ageofinformationvehicular}
S.~Kaul, M.~Gruteser, V.~Rai, and J.~Kenney, ``Minimizing age of information in
  vehicular networks,'' in \emph{Proc. of {IEEE SECON}}, Jun. 2011, pp.
  350--358.

\bibitem{kaul:ageofinformationqueue}
S.~Kaul, R.~Yates, and M.~Gruteser, ``Real-time status: How often should one
  update?'' in \emph{Proc. of {IEEE INFOCOM} Mini-Conference}, Mar. 2012, pp.
  2731--2735.

\bibitem{zinchenko:informationfreshness}
T.~Zinchenko, H.~Tchoankem, L.~Wolf, and A.~Leschke, ``Reliability analysis of
  vehicle-to-vehicle applications based on real world measurements,'' in
  \emph{Proc. of {ACM VANET} Workshop}, Jun. 2013, pp. 11--20.

\bibitem{tchouankem:messagelifetime}
H.~Tchoankem, T.~Zinchenko, and H.~Schumacher, ``Impact of buildings on
  vehicle-to-vehicle communication at urban intersections,'' in \emph{Proc. of
  {ICCC CCNC}}, Jan. 2015.

\bibitem{champati:ageofinformationfeedbackcontrol}
J.~P. Champati, M.~Mamduhi, K.~Johansson, and J.~Gross, ``Performance
  characterization using {AoI} in a single-loop networked control system,'' in
  \emph{Proc. of {IEEE INFOCOM} {AoI} Workshop}, Apr. 2019, pp. 197--203.

\bibitem{champati:ageofinformationmaxplus}
J.~P. Champati, H.~Al-Zubaidy, and J.~Gross, ``Statistical guarantee
  optimization for {AoI} in single-hop and two-hop {FCFS} systems with periodic
  arrivals,'' \emph{{IEEE} Trans. Commun.}, vol.~69, no.~1, pp. 365--381, Jan.
  2021.

\bibitem{altmann:ageofinformationmicroblogging}
E.~Altmann, R.~El-Azouzi, D.~S. Menasche, and Y.~Xu, ``Forever young: Aging
  control for hybrid networks,'' in \emph{Proc. of {ACM Mobihoc}}, Jul. 2019,
  pp. 91--100.

\bibitem{yates:ageofinformationsurvey}
R.~D. Yates, Y.~Sun, D.~R. Brown, S.~K. Kaul, E.~Modiano, and S.~Ulukus, ``Age
  of information: An introduction and survey,'' \emph{{IEEE} J. Sel. Areas
  Commun.}, vol.~39, no.~5, pp. 1183--1210, May 2021.

\bibitem{pappas:ageofinformationnetworkcalculus}
N.~Pappas and M.~Kountouris, ``Delay violation probabiliy and age of
  information interplay in the two-user multiple access channel,'' in
  \emph{Proc. of {IEEE SPAWC} Workshop}, Jul. 2019, pp. 1--5.

\bibitem{modiano:informationfreshness}
R.~Talak, S.~Karaman, and E.~Modiano, ``Optimizing information freshness in
  wireless networks under general interference constraints,'' \emph{{IEEE/ACM}
  Trans. Netw.}, vol.~28, no.~1, pp. 15--28, Feb. 2020.

\bibitem{modiano:ageofinformationqueueing}
L.~Huang and E.~Modiano, ``Optimizing age-of-information in a multi-class
  queueing system,'' in \emph{Proc. of {IEEE} International Symposium on
  Information Theory}, Jun. 2015, pp. 1681--1685.

\bibitem{champati:ageofinformationdgqueue}
J.~P. Champati, H.~Al-Zubaidy, and J.~Gross, ``Statistical guarantee
  optimization for age of information for the {D/G/1} queue,'' in \emph{Proc.
  of {IEEE INFOCOM} {AoI} Workshop}, Apr. 2018, pp. 130--135.

\bibitem{kam:ageofinformation}
C.~Kam, S.~Kompella, and A.~Ephremides, ``Age of information under random
  updates,'' in \emph{Proc. of {IEEE} International Symposium on Information
  Theory}, Jul. 2013, pp. 66--70.

\bibitem{champati:ageofinformationgigiqueue}
J.~P. Champati, H.~Al-Zubaidy, and J.~Gross, ``On the distribution of {AoI} for
  the {GI/GI/1/1} and {GI/GI/1/2*} systems: Exact expressions and bounds,'' in
  \emph{Proc. of {IEEE INFOCOM}}, Apr. 2019, pp. 37--45.

\bibitem{liebeherr:duality}
J.~Liebeherr, \emph{Duality of the Max-Plus and Min-Plus Network
  Calculus}.\hskip 1em plus 0.5em minus 0.4em\relax Now Publishers, 2017.

\bibitem{chang:dynamicserviceguarantees}
C.-S. Chang, R.~L. Cruz, J.-Y. {Le Boudec}, and P.~Thiran, ``A min, + system
  theory for constrained traffic regulation and dynamic service guarantees,''
  \emph{{IEEE/ACM} Trans. Netw.}, vol.~10, no.~6, pp. 805--817, 2002.

\bibitem{chang:performanceguarantees}
C.-S. Chang, \emph{Performance Guarantees in Communication Networks}.\hskip 1em
  plus 0.5em minus 0.4em\relax Springer-Verlag, 2000.

\bibitem{leboudec:networkcalculus}
J.-Y. {Le Boudec} and P.~Thiran, \emph{Network Calculus A Theory of
  Deterministic Queuing Systems for the {I}nternet}.\hskip 1em plus 0.5em minus
  0.4em\relax Springer-Verlag, 2001.

\bibitem{fidler:netcalcguide}
M.~Fidler and A.~Rizk, ``A guide to the stochastic network calculus,''
  \emph{{IEEE} Commun. Surveys Tuts.}, vol.~17, no.~1, pp. 92--105, Mar. 2015.

\bibitem{jiang:stochasticnetworkcalculus}
Y.~Jiang and Y.~Liu, \emph{Stochastic Network Calculus}.\hskip 1em plus 0.5em
  minus 0.4em\relax Springer-Verlag, Sep. 2008.

\bibitem{ciucu:networkservicecurvescaling2}
F.~Ciucu, A.~Burchard, and J.~Liebeherr, ``Scaling properties of statistical
  end-to-end bounds in the network calculus,'' \emph{{IEEE/ACM} Trans. Netw.},
  vol.~14, no.~6, pp. 2300--2312, Jun. 2006.

\bibitem{fidler:radiocalculus}
M.~Fidler, ``A network calculus approach to probabilistic quality of service
  analysis of fading channels,'' in \emph{{IEEE} {Globecom}}, Nov. 2006.

\bibitem{li:effectivebandwidthcalculus2}
C.~Li, A.~Burchard, and J.~Liebeherr, ``A network calculus with effective
  bandwidth,'' \emph{{IEEE/ACM} Trans. Netw.}, vol.~15, no.~6, pp. 1442--1453,
  Dec. 2007.

\bibitem{gibbens:markovonoff}
R.~J. Gibbens and P.~J. Hunt, ``Effective bandwidths for the multi-type {UAS}
  channel,'' \emph{Queueing Systems}, vol.~9, pp. 17--28, 1991.

\bibitem{kelly:effectivebandwidths}
F.~P. Kelly, ``Notes on effective bandwidths,'' ser. Royal Statistical Society
  Lecture Notes.\hskip 1em plus 0.5em minus 0.4em\relax Oxford University,
  1996, no.~4, pp. 141--168.

\end{thebibliography}
%
%
\end{document}